\documentclass[aps,pre,twocolumn,reprint,nofootinbib]{revtex4-2}
\pdfoutput=1

\usepackage{graphicx}
\usepackage{amssymb}
\usepackage{amsmath}
\usepackage{slashed}
\usepackage{color}
\usepackage{bbm}
\usepackage[T1]{fontenc} 

\definecolor{darkgreen}{rgb}{0.2,0.6,0}
\definecolor{lightblue}{rgb}{0,0.5,0.8}
\definecolor{lightred}{rgb}{0.8,0.2,0.2}
\definecolor{darkorange}{rgb}{1,0.549,0}

\usepackage[colorlinks, citecolor=darkgreen]{hyperref}

\newcommand{\BMW}{{\small BMW}}
\newcommand{\FRG}{{\small FRG}}
\newcommand{\eg}{{\textit{e.g.}}}
\newcommand{\ie}{{\textit{i.e.}}}

\allowdisplaybreaks

\begin{document}

\title{Exact solutions and residual regulator dependence \\ in functional renormalisation group flows}

\author{Benjamin Knorr\,\href{https://orcid.org/0000-0001-6700-6501}{\protect \includegraphics[scale=.07]{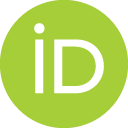}}\,}
\email[Electronic address: ]{bknorr@perimeterinstitute.ca}
\affiliation{Perimeter Institute for Theoretical Physics, 31 Caroline St N, Waterloo, ON N2L 2Y5, Canada}

\begin{abstract}
We construct exact solutions to the functional renormalisation group equation of the O(N) model and the Gross-Neveu model at large N for $2<d<4$, without specifying the form of the regulator. This allows to investigate which quantities are independent of the choice of regulator without being plagued by truncation artefacts. We find that only universal quantities, like critical exponents, and qualitative features, like the existence of a finite vacuum expectation value, are regulator-independent, whereas values of coupling constants are generically arbitrary. We also provide a general algorithm to construct a concrete operator basis for truncations in the derivative expansion and the Blaizot-M\'endez-Wschebor scheme.
\end{abstract}

\maketitle

\section{Introduction}\label{sec:Introduction}

The large $N$ limit is a useful and well-known expansion scheme in quantum field theory \cite{Moshe:2003xn}. By introducing an infinite number of a specific kind of field, one can control the leading order quantum effects, allowing for analytic expressions for, \eg{}, critical exponents characterising a second order phase transition as a series in $1/N$. In some cases, the (typically asymptotic) series can then be used to make predictions for a finite, physical number of fields. As a downside, it is generally hard to prove statements about the accuracy of such an expansion, and independent, non-perturbative computations are desirable to arrive at trustworthy estimates of physical quantities.

One such tool to perform non-perturbative calculations is the functional renormalisation group (\FRG{}) \cite{Wetterich:1992yh, Morris:1993qb, Berges:2000ew, Pawlowski:2005xe, Gies:2006wv, Delamotte:2007pf, Dupuis:2020fhh}. In this approach, one considers the so-called effective average action $\Gamma_k$, which depends on a fiducial momentum scale $k$. It interpolates between the classical action at a microscopic scale $\Lambda$, and the standard effective action at $k=0$. The $k$-dependence is dictated by the Wetterich equation \cite{Wetterich:1992yh, Morris:1993qb},
\begin{equation}\label{eq:FRGE}
 \dot \Gamma_k = \frac{1}{2} \text{Tr} \left[ \left( \Gamma_k^{(2)} + R_k \right)^{-1} \dot R_k \right] \, .
\end{equation}
In this, an overdot denotes the scale derivative $k \partial_k$, the trace includes a functional and an index trace as well as a minus sign for fermions, $\Gamma_k^{(2)}$ is the second variation of the effective average action, and $R_k$ is a regulator, which essentially acts as a $k$-dependent mass term,  making the right-hand side well-defined. In particular, in the limit $k\to0$, the regulator vanishes. This equation defines a vector field on theory space, and the points where this vector field vanishes are called fixed points. Fixed points correspond physically to second order phase transitions. The flow around such a fixed point is dictated by the universal critical exponents, which are related to the eigenvalues of the stability matrix at the fixed point.

For a general theory, the flow equation \eqref{eq:FRGE} typically has to be solved in an approximation\footnote{See however \cite{Ziebell:2020ckv} for recent developments in deriving general exact solutions to the flow equation.} which is not closed, \ie{}, in general one has to deal with systematic errors in the form of truncation errors. As a consequence, estimates of physical quantities obtained in truncated renormalisation group flows will rather generically show a residual regulator dependence \cite{Litim:2000ci, Litim:2001up, Pawlowski:2005xe, Pawlowski:2015mlf, Balog:2020fyt}. It is thus of interest to understand what quantities are truly independent of the choice of regulator in exact solutions to give a good hint for what kinds of regulator dependencies one should study in truncated renormalisation group flows. We will exemplify this by considering the large $N$ limit of some well-known theories: the $O(N)$ model and the (partially bosonised) Gross-Neveu model. These models are relevant to condensed matter and high energy physics, since they describe well-known second order phase transitions \cite{ZinnJustin:1989mi, Pelissetto:2000ek, Moshe:2003xn, Braun:2011pp, Gehring:2015vja}. As we will see, the large $N$ limit allows for a non-perturbative discussion of the existence and both universal and non-universal properties of the regular\footnote{Recently, non-regular solutions in this limit have been discussed in \cite{Yabunaka:2017uox, Katsis:2018bvc, Yabunaka:2018mju, Fleming:2020qqx}. These solutions are based on a more subtle large $N$ limit where naively sub-leading terms nevertheless contribute due to their divergence structure.} (Wilson-Fisher type) fixed point in these models. As a by-product, we will discuss some structural aspects of the flow equations in these models that carry over to finite $N$.

This work is structured as follows. In section \ref{sec:ONmodel} we discuss the $O(N)$ model, focussing on the large $N$ limit. We will first derive the simplified exact flow equation obtained in this limit in subsection \ref{sec:exactONflow}. In subsection \ref{sec:ONoperatorbasis}, we will construct an operator basis that parameterises the effective average action in the form of a derivative expansion, which reduces the functional flow equation to an infinite tower of first order ordinary differential equations. With that in place, in subsection \ref{sec:ONperfecttruncation} we show that the derivative expansion is a perfect truncation in the sense that the flow of any $n$-th order operator only depends on the solution of operators of lower or the same order. Since the effective potential plays a central role in the discussion of fixed points, we will investigate it in detail in subsection \ref{sec:ONpotential}. Subsection \ref{sec:ONresummation} is devoted to a partial resummation of the derivative expansion, which however still represents a perfect truncation. The resummation allows us to directly study momentum- and field-dependent correlation functions. With the complete fixed point solution, we are in the situation to discuss the regulator dependence in subsection \ref{sec:ONregdep}. Finally, we discuss some aspects of the model at finite $N$ in subsection \ref{sec:ONfiniteN}. As a second example, we discuss the Gross-Neveu model at large $N$ in section \ref{sec:GNmodel}. We first present the exact flow equation in the large $N$ limit in \ref{sec:GNflow}, then we discuss the lowest order critical correlation functions and critical exponents in \ref{sec:GNcorrelators} and \ref{sec:GNcritexp}, respectively, finishing the section with a short discussion of the regulator dependence in \ref{sec:GNregdep}. We close with a summary of the main results in section \ref{sec:summary}.

\section{O(N) model}\label{sec:ONmodel}
\subsection{Exact flow equation}\label{sec:exactONflow}

We will start our investigation by considering an $O(N)$-symmetric scalar field $\phi^a$ in $d$ Euclidean dimensions with $2<d<4$. Due to its simplicity and interesting applications, it is potentially the best-studied model in the context of the \FRG{} \cite{Benitez:2011xx, Codello:2014yfa, Defenu:2014bea, Borchardt:2015rxa, Hellwig:2015woa, Borchardt:2016pif, Borchardt:2016kco, Knorr:2016sfs, Litim:2016hlb, Eichhorn:2016hdi, Knorr:2017yze, Juttner:2017cpr, Defenu:2017dec, Marchais:2017jqc, Balog:2019rrg, DePolsi:2020pjk, Defenu:2020cji}. Let us derive the exact flow equation in the large $N$ limit for the effective action\footnote{To improve the readability, we will suppress the index $k$ on all quantities from hereon (except the regulator), and in a slight abuse of language refer to $\Gamma_k$ as the effective action. Since we will not discuss the limit $k\to0$ in this work, this should not lead to any confusion.}
\begin{equation}\label{eq:ONaction}
 \Gamma_{N\to\infty}[\phi] = \frac{N}{2} \int \text{d}^dx \, \left(\partial_\mu \phi^a\right) \left(\partial^\mu \phi^a\right) + N \bar\Gamma[\rho] \, ,
\end{equation}
where all non-trivial behaviour is carried by the second term $\bar\Gamma$. Here we introduced $\rho=\phi^a\phi^a/2$. The regulator is chosen to respect the $O(N)$ symmetry,
\begin{equation}
 \Delta S_k = \frac{N}{2} \int \text{d}^dx \, \phi^a R_k(-\partial^2) \phi^a \, .
\end{equation}
First we show that in the regular large $N$ limit the effective action is indeed of this form \eqref{eq:ONaction}. The crucial observation for the simplification in the large $N$ limit is that for the right-hand side of the flow equation to yield a factor of $N$, the trace over the $O(N)$ bundle indices must be a trace over the corresponding field space identity. For that reason, we only need the part of the two-point function which is proportional to the identity:
\begin{equation}
 \frac{\delta^2 \Gamma_{N\to\infty}}{\delta \phi^a \delta \phi^b} = N \left[ -\partial^2 + \frac{\delta\Gamma}{\delta\rho} \right] \delta^{ab} + \dots \, .
\end{equation}
Here, the dots indicate terms that, when traced over, do not give rise to a factor of $N$, \ie{}, their bundle indices are carried by $\phi$ and derivatives of it. Note that the identity component of the inverse of this operator only involves the inverse of the identity component of the operator itself. With this observation, we can immediately write down the \emph{exact} flow equation at large N:
\begin{equation}\label{eq:exact_flow_abstract}
 \dot{\bar\Gamma} = \frac{1}{2} \int \text{d}^dx \, \int \text{d}\mu_q \frac{\dot R_k(q^2)}{(q-\mathbf{i} \partial)^2 + R_k\left((q-\mathbf{i}\partial)^2\right) + \frac{\delta\bar\Gamma}{\delta\rho}} \, .
\end{equation}
In this we introduced the momentum integral measure
\begin{equation}
 \int \text{d}\mu_q = \int \frac{\text{d}^dq}{(2\pi)^d} \, .
\end{equation}
Two points deserve being mentioned with regards to \eqref{eq:exact_flow_abstract}: first, the full trace is already performed, and only a standard momentum integral is left. Second, in contrast to standard flows, only the \emph{first} functional derivative of the action appears in the propagator on the right-hand side. This is indeed at the heart of the simplifications in the large $N$ limit, and allows to write down perfect truncations where increasing the truncation order doesn't alter the flow equations of previously calculated orders.

We still have to argue that the standard kinetic term used above is the only possibility for a term which doesn't depend on $\rho$ only. In particular, one could imagine a field- or momentum-dependent wave function renormalisation factor, also giving rise to a potentially non-vanishing anomalous dimension. However, since the flow equation doesn't give any contribution to the flow of the kinetic term, requiring locality and regularity at vanishing field immediately fixes a trivial wave function renormalisation and a vanishing anomalous dimension.

\subsection{Spanning an operator basis}\label{sec:ONoperatorbasis}

Even though the flow equation \eqref{eq:exact_flow_abstract} is much simpler than it is in the generic case, a direction solution is still involved. To make things tractable, we will now discuss spanning the non-trivial part of the effective action, $\bar\Gamma$, in a derivative expansion. For this, we need to specify a unique basis at every fixed number of derivatives where we took partial integrations into account. Since derivatives always have to come in pairs to form a Lorentz scalar, we will refer to the $n$-th order as all terms having $2n$ derivatives. The convergence properties of this expansion at finite $N$ have been thoroughly discussed in \cite{Balog:2019rrg, DePolsi:2020pjk}.

\begin{figure}[t]
 \includegraphics[width=0.4\columnwidth]{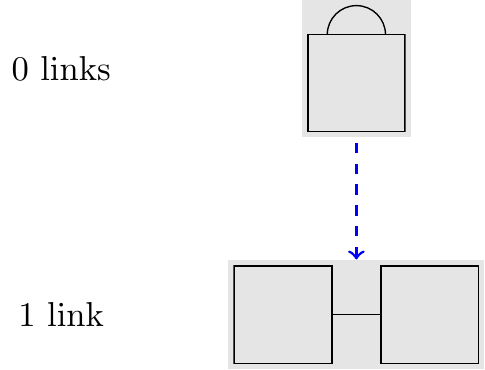}
 \caption{The graph indicating the partial integration rules and the basis for the first order of the derivative expansion for the $O(N)$ model at large $N$. The pictograph in the first line correspond to an operator $X(\rho) \partial^2 \rho$, which is related to the operator represented by the pictograph in the second line, $Y(\rho) (\partial_\mu \rho) (\partial^\mu \rho)$, by a partial integration. The blue dashed line indicates that $X$ and $Y$ are related by $Y(\rho) = -X'(\rho)$.}
 \label{fig:d2_graph}
\end{figure}

At a given order $n$, a general operator in the derivative expansion of $\bar\Gamma$ has the form
\begin{equation}\label{eq:op_form}
 X(\rho) \left( \partial_{\mu} \cdots \rho \right) \left( \partial_\nu \cdots \rho \right) \cdots \, ,
\end{equation}
that is, it is a product of a $\rho$-dependent function and a series of derivatives acting on different $\rho$s so that for constant $\rho$, all interaction terms except for the potential vanish. We will now describe a graphical procedure to obtain a unique basis at order $n$, and all partial integration rules to write an arbitrary operator in this basis. For this, we draw a box for every $\rho$ on which derivatives act, and a link connecting two boxes (or a box with itself) which corresponds to a pair of derivatives which is contracted, the end points marking the fields on which they act. With this, the algorithm at order $n$ is as follows:
\begin{enumerate}
 \item In the first row, draw all possible graphs with any number of boxes and exactly $n$ links, where all links start and end on the same box, and no box can have zero links attached to it. This corresponds to all operators of the form \eqref{eq:op_form} where we only have $\partial^2$ as derivative operators, each acting on a single $\rho$.
 \item To arrive at the next row of the graph, employ the following partial integration rules. For any graph of the first row, from the box with the most links, disconnect one of the links which starts and ends on the box, and draw all distinct graphs where the loose link is connected to either one of the existing boxes, or a new box, with the following weights:
  \begin{enumerate}
   \item if the link is connected to an already existing box, multiply with a factor of $-1$,
   \item if the link is connected to a new box, multiply with a factor of $-x$.
  \end{enumerate}
 If connecting the loose link to different boxes gives the same new diagram, multiply the weight by this multiplicity.
 \item Draw lines from every ``parent'' box of the given level of the graph to its ``children'' as obtained by the previous integration rules, indicating the weights.
 \item Perform the two previous steps for all other links, but don't attach weights to the lines from the parents to these ``illegitimate children''. This step ensures that all distinct graphs on a given row are generated, since not all graphs of a given row need to have parents in the above sense. The first time that new operators are generated in this way for the $O(N)$ model is at fourth order.
 \item Repeat the above steps until all links connect different boxes.
\end{enumerate}
The bottom line of the graph represents the independent basis at order $n$ that we will use, and corresponds to the basis where all $\partial^2$ operators are partially integrated. To write any element in this basis, follow the graph starting from this element down to the bottom row, multiplying the weights from row to row. Once the bottom row is reached, any factor of $x$ corresponds to a $\rho$-derivative of the operators' prefactor, that is the function $X$ in \eqref{eq:op_form}. To illustrate this procedure, we show the graphs for orders $n=1,2,3$ in \autoref{fig:d2_graph}, \autoref{fig:d4_graph} and \autoref{fig:d6_graph}, the graph for $n=4$ can be found in the supplemental material as it is too bulky to present it here. In these graphs, we only indicate multiplicities explicitly. Factors of $-1$ are indicated by olive, full lines while factors of $-x$ are indicated by blue, dashed lines. Illegitimate relations are indicated by red, dotted lines.

We can now span the effective action in terms of our basis in a derivative expansion:
\begin{equation}
\begin{aligned}
 \bar\Gamma &= \int\text{d}^dx \, \Bigg[ V(\rho) + Y(\rho) \left(\partial_\mu \rho \right) \left(\partial^\mu \rho \right) \\
 &+ W_1(\rho) \left( \partial_\mu \partial_\nu \rho \right) \left( \partial^\mu \partial^\nu \rho \right) + W_2(\rho) \left( \partial_\mu \rho \right) \left( \partial^\mu \partial^\nu \rho \right) \left( \partial_\nu \rho \right) \\
 &+ W_3(\rho) \left(\partial_\mu \rho \right) \left(\partial^\mu \rho \right) \left(\partial_\nu \rho \right) \left(\partial^\nu \rho \right) + \mathcal O(\partial^6) \Bigg] \, .
\end{aligned}
\end{equation}
The potential $V$ has no graphical representation. The wave function renormalisation for the field $\rho$ is represented in the bottom line of \autoref{fig:d2_graph}, and the functions $W_{1,2,3}$ correspond to the bottom line of \autoref{fig:d4_graph}, from left to right. The number of terms increases rapidly with the order -- for $n=3$, there are eight coupling functions, whereas for $n=4$, there are already 23 coupling functions.

\begin{figure}[t!]
 \includegraphics[width=\columnwidth]{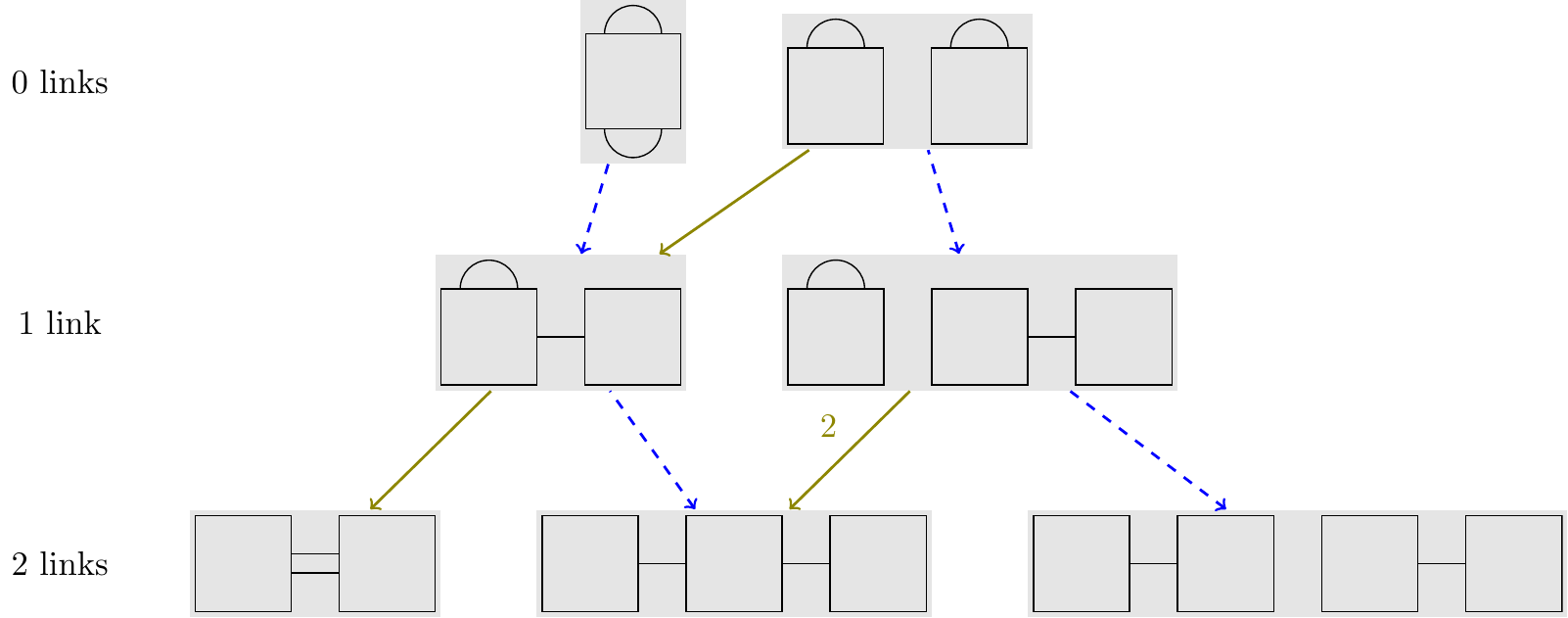}
 \caption{The graph indicating the partial integration rules and the basis for the second order of the derivative expansion for the $O(N)$ model at large $N$.}
 \label{fig:d4_graph}
\end{figure}

\begin{figure*}
 \includegraphics[width=\textwidth]{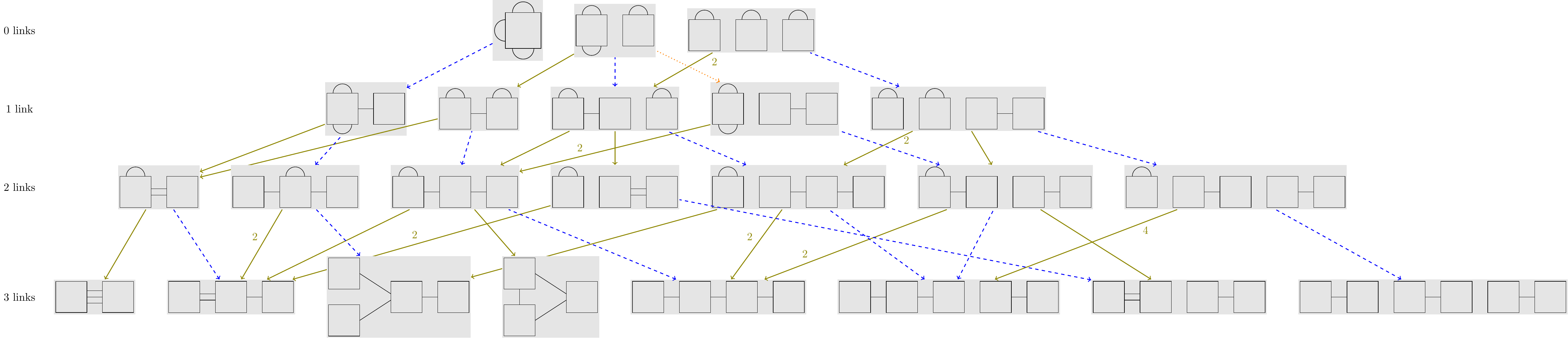}
 \caption{The graph indicating the partial integration rules and the basis for the third order of the derivative expansion for the $O(N)$ model at large $N$.}
 \label{fig:d6_graph}
\end{figure*}

\subsection{Derivative expansion as a perfect truncation}\label{sec:ONperfecttruncation}

In this section we will show that the derivative expansion in the large $N$ limit is a perfect truncation, that is, the flow of the $n$-th order coupling functions doesn't depend on the coupling functions of any higher order. In fact, we can show that for every coupling function $X$ except the potential, the flow equation reads
\begin{equation}\label{eq:flowgenericX}
\begin{aligned}
 \dot X(\rho) &= -\frac{1}{2} X'(\rho) \int \text{d}\mu_q \frac{\dot R_k(q^2)}{\left(q^2+R_k(q^2)+V'(\rho)\right)^2} \\
 &+ n_X V''(\rho) X(\rho) \int \text{d}\mu_q \frac{\dot R_k(q^2)}{\left(q^2+R_k(q^2)+V'(\rho)\right)^3} \\
 &+ \mathfrak f_X(\rho) \, ,
\end{aligned}
\end{equation}
where $n_X$ is the number of $\rho$s in the operator on which derivatives act. Graphically, this is the number of boxes in the graph associated to the coupling function $X$. Furthermore, $\mathfrak f_X(\rho)$ is a function which doesn't depend on $X$, but depends on some operators of lower order, and can depend linearly on coupling functions of the same order, but not their derivatives. Let us point out that the flow equations are thus first order linear ordinary differential equations.

To show this form of the flow equation, we will use the basis introduced in the previous section. For a general operator that appears in the action as
\begin{equation}
 X(\rho) \left( \mathcal D_1 \rho \right) \cdots \left( \mathcal D_{n_X} \rho \right) \, ,
\end{equation}
where the $\mathcal D_i$ are collections of contracted derivatives in the form of the basis of the previous section, the variation w.r.t.\ $\rho$ reads
\begin{widetext}
\begin{equation}
\begin{aligned}
 &\frac{\delta}{\delta \rho} \int \text{d}^dx \, X(\rho) \left( \mathcal D_1 \rho \right) \cdots \left( \mathcal D_{n_X} \rho \right) \\
 &= X'(\rho) \left( \mathcal D_1 \rho \right) \cdots \left( \mathcal D_{n_X} \rho \right) + \sum_{l=1}^{n_X} \left[ (-1)^{|\mathcal D_l|} \mathcal D_l \left( X(\rho) \left(\mathcal D_1 \rho \right) \cdots \left( \mathcal D_{l-1}\rho \right) \left( \mathcal D_{l+1}\rho \right) \cdots \left( \mathcal D_{n_X} \rho \right) \right) \right] \, .
\end{aligned}
\end{equation}
\end{widetext}
In this, $|\mathcal D_l|$ is the number of derivatives contained in the operator $\mathcal D_l$. Since this expression is still of the same derivative order, it cannot contribute to the flow of operators of lower orders, and we can expand the flow equation to linear order in this expression to obtain its contribution to the same order. The first term is still in the correct basis, so that it only contributes to the flow of itself at that order, and is indeed the first term of \eqref{eq:flowgenericX}. The additional minus sign comes from the expansion of the propagator. The second term gives a contribution
\begin{widetext}
\begin{equation}
 -\frac{1}{2} \int \text{d}^dx \, \int \text{d}\mu_q \frac{\dot R_k(q^2)}{\left( q^2 + R_k(q^2) + V'(\rho) \right)^2} \sum_{l=1}^{n_X} \left[ (-1)^{|\mathcal D_l|} \mathcal D_l \left( X(\rho) \left(\mathcal D_1 \rho \right) \cdots \left( \mathcal D_{l-1}\rho \right) \left( \mathcal D_{l+1}\rho \right) \cdots \left( \mathcal D_{n_X} \rho \right) \right) \right]
\end{equation}
\end{widetext}
We have to perform a partial integration on $\mathcal D_l$ to arrive again in our chosen basis. Since the $\mathcal D_l$ don't include $\partial^2$ by construction, performing the partial integration directly gives an expression in the correct basis. In particular, we can easily isolate the term which contributes to the flow of $X$ itself: it is that term where all the derivatives of $\mathcal D_l$ act on a single $\rho$. From this, we get a factor of $-2V''(\rho)$ and another propagator from taking the derivative of the squared propagator, and a factor of $n_X$ from the sum, so that we arrive at the second line of \eqref{eq:flowgenericX}. All other terms that $X$ contributes to are different operators of the same order, where it contributes linearly and without derivatives, or higher orders. This completes the proof of \eqref{eq:flowgenericX}.

\subsection{Fixed point potential}\label{sec:ONpotential}

We will now start to discuss the fixed point, that is the point where the flow of all dimensionless coupling functions vanishes. For this, we rescale the field and coupling functions by appropriate powers of $k$, \eg{},
\begin{equation}
 \rho = k^{d-2} r \, , \qquad V(\rho) = k^d v( k^{2-d} \rho) \, .
\end{equation}
Here, $r$ is the dimensionless version of the field $\rho$ and $v$ the dimensionless potential.\footnote{Since the wave function renormalisation factor $Z$ doesn't renormalise in the large $N$ limit of the $O(N)$ model, we don't have to include it in the rescaling. This is different in the Gross-Neveu model discussed below, where we assume that proper powers of this factor have been included in all dimensionless couplings.} With this, at the fixed point we have to solve
\begin{equation}\label{eq:flowv}
 d \, v(r) - (d-2) \, r \, v'(r) = \frac{1}{2} \int \text{d}\mu_q \frac{\dot R_k(q^2)}{q^2+R_k(q^2) + v'(r)} \, ,
\end{equation}
to arrive at the critical potential. In fact, it is easier to solve the derivative of this equation for $u(r)\equiv v'(r)$:
\begin{equation}\label{eq:flowu}
\begin{aligned}
 &2u(r) - (d-2) \, r \, u'(r) \\
 & = -\frac{1}{2} u'(r) \int \text{d}\mu_q \frac{\dot R_k(q^2)}{\left(q^2+R_k(q^2) + u(r)\right)^2} \, .
\end{aligned}
\end{equation}
The Wilson-Fisher type solution of this equation can be given implicitly,
\begin{equation}\label{eq:crit_pot}
\begin{aligned}
 r &= \frac{1}{2(d-2)} \int \text{d}\mu_q \frac{\dot R_k(q^2)}{\left( q^2 + R_k(q^2) \right)^2} \times \\
 &\qquad {}_2F_1\left( 2, 1-\frac{d}{2}; 2-\frac{d}{2} \bigg| -\frac{u(r)}{q^2+R_k(q^2)} \right) \, .
\end{aligned}
\end{equation}
The complete set of solutions of \eqref{eq:flowu} for a specific choice of regulator is discussed, \eg{}, in \cite{Marchais:2017jqc}.

In fact, \eqref{eq:crit_pot} establishes a bijective map between $r\in(-\infty,\infty)$ and $u\in(u_{-\infty},\infty)$, even though a priori physical values of $r$ are non-negative. To see this, we calculate $u'(r)$ and show that it is positive. Taking the derivative of \eqref{eq:crit_pot} with respect to $r$ and solving for $u'$, we get
\begin{widetext}
\begin{equation}\label{eq:uprimeofu}
 u'(r) = \left[ \frac{1}{4} \int \text{d}\mu_q \frac{\dot R_k(q^2)}{\left(q^2+R_k(q^2)\right)^3} \frac{q^2+R_k(q^2)}{u(r)} \left\{ {}_2F_1\left( 2, 1-\frac{d}{2}; 2-\frac{d}{2} \bigg| -\frac{u(r)}{q^2+R_k(q^2)} \right) - \left( \frac{1}{1+\frac{u(r)}{q^2+R_k(q^2)}} \right)^2 \right\} \right]^{-1} \, .
\end{equation}
\end{widetext}
The right-hand side is positive as long as
\begin{equation}
 u(r)>-\min_q(q^2+R_k(q^2)) \equiv u_{-\infty} \, .
\end{equation}
In particular, for any regulator $u_{-\infty}$ is negative, since from general regulator criteria $q^2 + R_k(q^2)>0$. It is easy to see that $u=u_{-\infty}$ corresponds to $r=-\infty$ due to the properties of the hypergeometric function in \eqref{eq:crit_pot}.

Independent of the regulator, this establishes that there must be a finite $r_0$ such that $u$ vanishes, that is the potential has a minimum. Inserting $u(r_0)=0$ into the implicit solution gives the explicit expression
\begin{equation}\label{eq:ONvev}
 r_0 = \frac{1}{2(d-2)} \int \text{d}\mu_q \frac{\dot R_k(q^2)}{\left( q^2 + R_k(q^2) \right)^2} \, ,
\end{equation}
which is positive for any regulator, so that the minimum is always in the physical regime.

\begin{figure}
 \includegraphics[width=0.9\columnwidth]{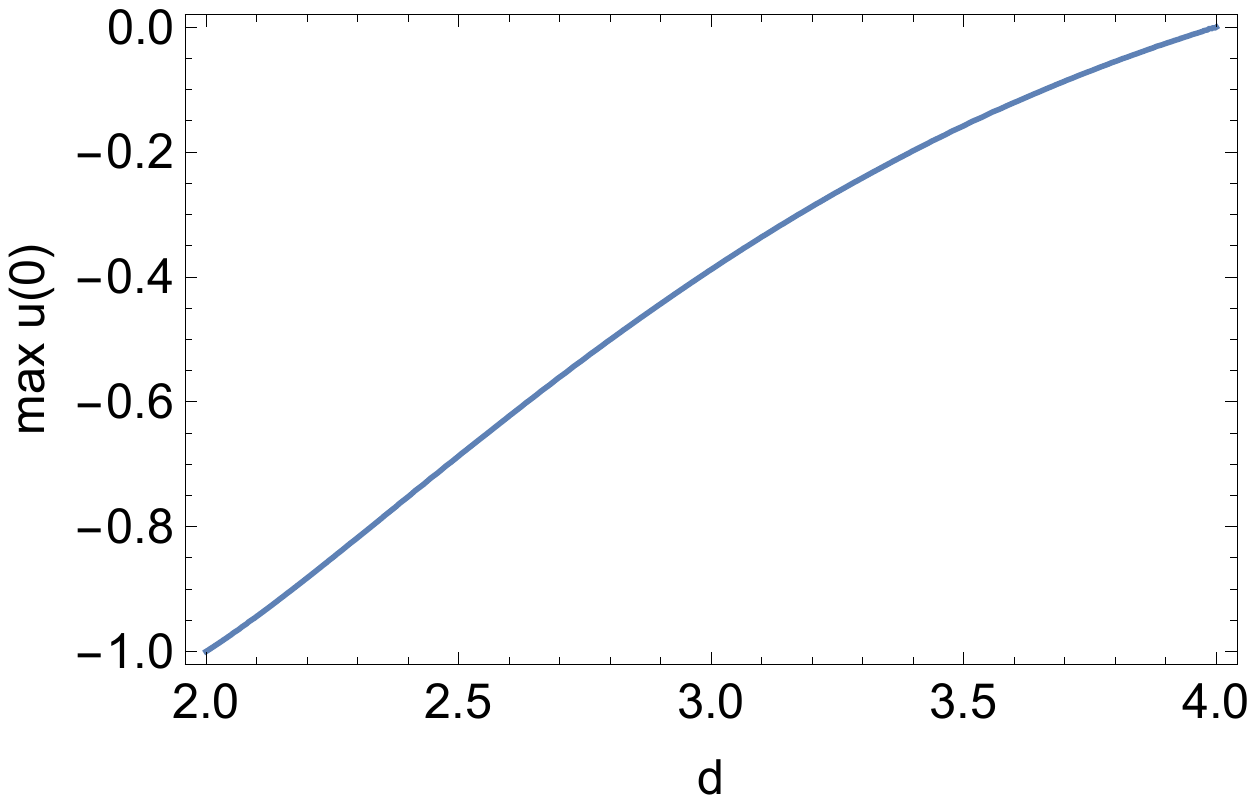}
 \caption{The maximal allowed value of $u(0)$ for dimensions $2<d<4$. This universal bound must be fulfilled for any regulator choice. The bound is saturated for the Litim regulator \cite{Litim:2001up}.}
 \label{fig:maxu0}
\end{figure}

By evaluating \eqref{eq:crit_pot} at $r=0$, we get an implicit equation which fixes $u(0)$. This indeed also gives an upper bound on $u(0)$ since for the integral to vanish, the argument of the hypergeometric function has to have a minimum value, as it is positive for all arguments $x$ larger than some small negative value. The bound for $u(0)$ is shown in \autoref{fig:maxu0} for general dimensions $d$. In $d=3$, the hypergeometric function vanishes at approximately $0.388$. This extremal value is exactly reached for the Litim cutoff \cite{Litim:2001up}, for which the hypergeometric function is independent of the integration variable and thus its argument must be its zero.

The bijective relation between $r$ and $u$ will play a central role in the following discussion, since as it turns out, $u$ is a much more natural variable than $r$.

At this point it is also worthwhile to note the special form of the fixed point equations for all the higher order operators. Translating \eqref{eq:flowgenericX} into dimensionless form, we have
\begin{equation}
\begin{aligned}
 &d_x \, x(r) - (d-2) \, r \, x'(r) \\
 &= -\frac{1}{2} x'(r) \int \text{d}\mu_q \frac{\dot R_k(q^2)}{\left(q^2+R_k(q^2)+u(r)\right)^2} \\
 &+ n_X u'(r) x(r) \int \text{d}\mu_q \frac{\dot R_k(q^2)}{\left(q^2+R_k(q^2)+u(r)\right)^3} + \mathfrak f_X(r) \, ,
\end{aligned}
\end{equation}
where $d_x$ and $x$ are the canonical dimension and the dimensionless version of the coupling function $X$, respectively. Using \eqref{eq:flowu} to replace the first integral on the right-hand side gives
\begin{equation}
\begin{aligned}
 &d_x \, x(r) - 2\frac{u(r)}{u'(r)} x'(r) \\
 &= n_X u'(r) x(r) \int \text{d}\mu_q \frac{\dot R_k(q^2)}{\left(q^2+R_k(q^2)+u(r)\right)^3} + \mathfrak f_X(r) \, .
\end{aligned}
\end{equation}
This is a linear first order differential equation for $x$, so a priori it has one free integration constant. We notice however that at $r=r_0$ the prefactor of $x'$ vanishes, which fixes the integration constant uniquely by requiring that the solution is regular.\footnote{Strictly speaking, since $\mathfrak f_X$ contains coupling functions of the same order, one has to deal with coupled linear equations. Showing existence of a solution is easier for the resummed correlation functions discussed below, so we will not dwell on it here.}

Before we go on with the discussion of the other operators, let us recalculate the critical exponents, showing that they are indeed regulator-independent. Perturbing the flow equation for the potential linearly about the fixed point gives
\begin{equation}
\begin{aligned}
 &(d-\theta) \delta v(r) - (d-2) \, r \, \delta v'(r) \\
 &= -\frac{1}{2} \delta v'(r) \int \text{d}\mu_q \frac{\dot R_k(q^2)}{\left(q^2+R_k(q^2) + v'(r)\right)^2} \\
 &= \left( 2 \frac{v'(r)}{v''(r)} - (d-2) \, r \right) \delta v'(r) \, ,
\end{aligned}
\end{equation}
where $\delta v$ is the perturbation of the fixed point potential $v$ and $\theta$ is the critical exponent. In the last line, we used the fixed point equation \eqref{eq:flowu} to replace the integral. We can now use the aforementioned variable transformation to map $r$ to $u$ and get
\begin{equation}
 (d-\theta) \delta v(u) - 2 u \, \delta v'(u) = 0 \, ,
\end{equation}
which has the solution
\begin{equation}
 \delta v(u) = c \, u^{\frac{d-\theta}{2}} \, ,
\end{equation}
with $c$ being a normalisation constant. Since $u(r_0)=0$, and $u(r)<0$ for $0 \leq r<r_0$, we have to demand that the exponent is a non-negative integer $n$ for the perturbations to be globally well-defined and real, which gives the well-known quantised spectrum
\begin{equation}\label{eq:ONcritexp}
 \theta = d-2n \, , \qquad n \in \mathbbm N \, .
\end{equation}
To complete the argument, we still have to show that there are perturbations for all higher order operators such that they fulfil their perturbed flow equations for the same critical exponents, this however follows from similar arguments as the existence of the fixed point itself.

\subsection{Partial resummation of the derivative expansion}\label{sec:ONresummation}

With the results provided so far, in principle the full fixed point effective action can be reconstructed, order by order in the derivative expansion. As it turns out, this procedure can be optimised by resumming certain subclasses of operators. This comes from the observation that an operator corresponding to a diagram element with $n$ boxes only contributes to diagram elements with $n$ or more boxes, that is, there is a further type of hierarchy which gives again rise to a perfect truncation. To prove this statement, we use momentum space techniques, and consider the $n$-th $\rho$-derivative of the flow equation, evaluated at constant $\rho$. Because in the large $N$ limit, the propagator only involves the first derivative of the effective action, the flow of the $n$-point function involves correlators up to order $n+1$. In particular, the only dependence on the higher order correlator is by a tadpole diagram. Explicitly, this dependence is
\begin{equation}
\begin{aligned}
 &\dot \Gamma^{(n)}(p_1,\dots,p_n | \rho) \supset -\frac{1}{2} \int \text{d}\mu_q \frac{\dot R_k(q^2)}{\left(q^2+R_k(q^2)+V'(\rho)\right)^2} \times \\
 & \qquad\qquad\qquad\qquad \Gamma^{(n+1)}(p_1, \dots, p_n, -(p_1+\dots+p_n) | \rho) \, .
\end{aligned}
\end{equation}
For better readability, we introduced a vertical bar to separate the momentum arguments from the field argument. The dependence of the $n$-point functions $\Gamma^{(n)}$ on the momenta $p_i$ is fully symmetric. Using momentum conservation for the last momentum argument of the $(n+1)$-point function, we find
\begin{equation}
\begin{aligned}
 &\dot \Gamma^{(n)}(p_1,\dots,p_n) | \rho)
 \\ &\supset -\frac{1}{2} \Gamma^{(n+1)}(p_1, \dots, p_n, 0 | \rho) \int \text{d}\mu_q \tfrac{\dot R_k(q^2)}{\left(q^2+R_k(q^2)+V'(\rho)\right)^2} \, .
\end{aligned}
\end{equation}
The key observation is that the last momentum in the $(n+1)$-point function is zero, so that there is no contribution from operators with $(n+1)$ boxes.

Readers familiar with \FRG{} techniques will realise that this partial resummation is a concrete realisation of the so-called Blaizot-M\'endez-Wschebor (\BMW{}) approximation \cite{Blaizot:2005xy, Blaizot:2005wd, Blaizot:2006vr, Benitez:2009xg, Benitez:2011xx}, which resolves both momentum and field dependence. In fact, the diagrammatic representation can be used as a way to give a concrete action as a starting point to the \BMW{} scheme, and motivates the origin of the truncation rules employed there. Concretely, the resummed ansatz reads in our case
\begin{widetext}
\begin{equation}\label{eq:BMW-exp}
\begin{aligned}
 &\bar\Gamma = \int \text{d}^dx \, \Bigg[ V(\rho) + \sum_{l\geq1} Y_l(\rho) \left( \partial_{\mu_1} \cdots \partial_{\mu_l} \rho \right) \left( \partial^{\mu_1} \cdots \partial^{\mu_l} \rho \right) \\
 &+ \sum_{n\geq2}\sum_{l=1}^{n-1} W_{-,n,l}(\rho) \left( \partial_{\mu_1} \cdots \partial_{\mu_l} \rho \right) \left( \partial^{\mu_1} \cdots \partial^{\mu_l} \partial_{\mu_{l+1}} \cdots \partial_{\mu_n} \rho \right) \left( \partial^{\mu_{l+1}} \cdots \partial^{\mu_n} \rho \right) \\
 &+ \sum_{n\geq3} \sum_{l=1}^{\lfloor\frac{n}{3}\rfloor} \sum_{j=l}^{\lfloor \frac{n-j}{2} \rfloor} W_{\Delta,n,l,j}(\rho) \left( \partial_{\mu_1} \cdots \partial_{\mu_l} \partial_{\mu_{l+1}} \cdots \partial_{\mu_{l+j}} \rho \right) \left( \partial^{\mu_{l+1}} \cdots \partial^{\mu_{l+j}} \partial^{\mu_{l+j+1}} \cdots \partial^{\mu_n} \rho \right) \left( \partial^{\mu_1} \cdots \partial^{\mu_l} \partial_{\mu_{l+j+1}} \cdots \partial_{\mu_n} \rho \right) \\
 &+ \mathcal O \left((\partial \rho)^4\right) \Bigg] \, .
\end{aligned}
\end{equation}
\end{widetext}
The first line corresponds graphically to $0$ and $2$ boxes, the second line comprises the diagrams with three boxes arranged in a line, whereas the third line lists all operators with three boxes arranged in a triangle. To our knowledge, such an explicit expression for the effective action giving rise to the \BMW{} scheme hasn't been put forward before.

We will now iteratively study the flow equations for the $n$-point functions to gain some structural insights. The zero-point function is identical to the potential and has been dealt with in the last subsection. The first non-trivial momentum dependence arises for the two-point function.\footnote{We emphasise that the two-point function of the composite $\rho$ corresponds to a combination of parts of the two-, three- and four-point functions of the elementary field $\phi$. Similarly, higher order composite correlators correspond to combinations of higher order elementary correlators.}

\subsubsection{Two-point function}\label{sec:ONtwopoint}

Taking the second $\rho$-derivative of the flow equation and evaluating it for constant field gives
\begin{widetext}
\begin{equation}
\begin{aligned}
 \dot{\bar \Gamma}^{(0,2)}(p^2|\rho) &= -\frac{1}{2} \frac{\delta^3 \bar\Gamma}{\delta \rho(p) \delta \rho(-p) \delta \rho(0)} \int \text{d}\mu_q \frac{\dot R_k(q^2)}{\left(q^2+R_k(q^2)+V'(\rho)\right)^2} \\
 &\qquad+ \left[\bar\Gamma^{(0,2)}(p^2|\rho)\right]^2 \int \text{d}\mu_q \frac{\dot R_k(q^2)}{\left(q^2+R_k(q^2)+V'(\rho)\right)^2} \frac{1}{(p+q)^2 + R_k((p+q)^2) + V'(\rho)} \, .
\end{aligned}
\end{equation}
\end{widetext}
Note that no loop momentum appears in the vertex functions. This is a general feature in the large $N$ limit and once again arises due to the fact that the exact flow equation \eqref{eq:exact_flow_abstract} depends on the first, not the second derivative of the effective action. For that reason, the two internal $\phi$-legs of any vertex actually count as a single $\rho$-leg, with a momentum equal to minus all external legs due to momentum conservation, and hence no loop momentum. As a consequence, we still only have to deal with (now partial) differential equations and not integro-differential equations, as would be expected when resolving momentum dependencies.

We will call the dimensionless version of the two-point vertex $\gamma_2$. Furthermore we will introduce the integrals
\begin{align}
 &\mathcal I_n(u(r)) = \int \text{d}\mu_q \frac{\dot R_k(q^2)}{\left(q^2+R_k(q^2)+u(r)\right)^n} \, , \\
 &\mathcal I_{n,j}(p_1, \dots, p_j|u(r)) = \int \text{d}\mu_q \frac{\dot R_k(q^2)}{\left(q^2+R_k(q^2)+u(r)\right)^n} \times \notag \\
 & \qquad\qquad \prod_{l=1}^j \tfrac{1}{(p_1+\cdots+p_l+q)^2 + R_k((p_1+\cdots+p_l+q)^2)+u(r)} \, .
\end{align}
Taking the above result that the tadpole contribution to the flow doesn't include any higher order correlators, we find that the fixed point equation for $\gamma_2$ reads
\begin{equation}
\begin{aligned}
 &(4-d) \gamma_2(p^2|r) - (d-2) \, r \, \gamma_2^{(0,1)}(p^2|r) - 2 p^2 \, \gamma_2^{(1,0)}(p^2|r) \\
 &= - \frac{1}{2} \gamma_2^{(0,1)}(p^2|r) \mathcal I_2(u(r)) + \gamma_2(p^2|r)^2 \mathcal I_{2,1}(p^2|u(r)) \, .
\end{aligned}
\end{equation}
Note that we made it explicit in the argument of $\mathcal I_{2,1}$ that because of Lorentz invariance, the integral must be a function of the squared dimensionless momentum, which we denote by the same letter $p^2$. We can simplify the equation in two steps. First, we again use the fixed point equation for the derivative of the potential to replace the first integral on the right-hand side of the equation via \eqref{eq:flowu}. Second, we use the bijective map from $r$ to $u$ and make a coordinate transformation to replace all occurrences of $r$ by u. This results in the fixed point equation
\begin{equation}
\begin{aligned}
 &(4-d) \gamma_2(p^2|u) - 2p^2 \gamma_2^{(1,0)}(p^2|u) - 2u \gamma_2^{(0,1)}(p^2|u) \\
 &= \gamma_2(p^2|u)^2 \, \mathcal I_{2,1}(p^2|u) \, .
\end{aligned}
\end{equation}
Taking care of the correct regular boundary conditions, this gives rise to the solution
\begin{equation}
 \gamma_2(p^2|u) = \frac{1}{\frac{1}{2} \int_0^1 \text{d}\omega \, \omega^{1-\frac{d}{2}} \mathcal I_{2,1}(\omega \, p^2 | \omega \, u)} \, .
\end{equation}
The integral exists for all $d\in(2,4)$. It can be checked straightforwardly by comparison with \eqref{eq:uprimeofu} and using identities for ${}_2F_1$ that for $p^2=0$ indeed $\gamma_2(0,u) = u'(r)$, as it should be. Expanding the solution in a Taylor series in $x$ around $0$, one can directly read off the (dimensionless) fixed point coefficient functions $y_l(r)$ of the ansatz \eqref{eq:BMW-exp}. Explicitly,
\begin{equation}
 \gamma_2(p^2|u) = u'(r) + 2 \sum_{l\geq 1} y_l(r) p^{2l} \, ,
\end{equation}
where both $r$ and $u'(r)$ have to be replaced according to \eqref{eq:crit_pot} and \eqref{eq:uprimeofu}.

We can also derive the large momentum limit at finite $u$. In this limit,
\begin{equation}
 \gamma_2(p^2|u) \propto \left( p^2 \right)^{2-\frac{d}{2}} \, , \qquad \text{as } p \to \infty \, .
\end{equation}
Comparing to the standard behaviour of a two-point function at large momentum,
\begin{equation}
 \gamma_2(p^2|u) \propto \left( p^2 \right)^{1-\frac{\eta}{2}} \, ,
\end{equation}
this suggests to define an anomalous dimension with respect to the field $\rho$ with value
\begin{equation}
 \eta_\rho = d-2 \, .
\end{equation}
This is to be contrasted with the anomalous dimension of the fundamental field $\phi$, which vanishes. Note also that while the power law exponent is regulator-independent, the overall prefactor of the limit does depend on the choice of the regulator.

\subsubsection{Higher order correlation function}\label{sec:ONnpoint}

For the higher order correlation functions, it is convenient to choose the inner product of distinct momenta as a basis for the momentum dependence,
\begin{equation}
 y_{ij} = p_{i\mu} p_j^\mu \, , \qquad 1 \leq i < j \leq n \, .
\end{equation}
Any squared momentum can be replaced by a combination of these by imposing momentum conservation,
\begin{equation}
 \sum_{l=1}^n p_l = 0 \, ,
\end{equation}
where we have the convention that all momenta are ingoing. This implies the replacement rules
\begin{equation}
 p_i^2 = - p_{i\mu} \sum_{\substack{l=1 \\ l\neq i}}^n p_l^\mu  = -\sum_{l=1}^{i-1} y_{li} -\sum_{l=i+1}^n y_{il} \, .
\end{equation}
Whenever squared momenta appear in the following, they are to be understood as a shorthand for their expression in terms of the $y_{ij}$.

Following the same steps as for the two-point function, we get the fixed point equation for the three-point function,
\begin{equation}
\begin{aligned}
 &\left[(6-2d) - 2y_{ij} \, \partial_{y_{ij}} - 2u \, \partial_u \right] \gamma_3(y_{12},y_{13},y_{23}|u) \\
 &= \gamma_3(y_{12},y_{13},y_{23}|u) \sum_{i=1}^3 \gamma_2(p_i^2|u) \mathcal I_{2,1}(p_i^2|u) \\
 &-\frac{1}{2} \gamma_2(p_1^2|u) \gamma_2(p_2^2|u) \gamma_2(p_3^2|u) \sum_{\substack{i,j=1 \\ i\neq j}}^3 \mathcal I_{2,2}(p_i,p_j|u) \, .
\end{aligned}
\end{equation}
By requiring momentum conservation and again using Lorentz symmetry, it is clear that all terms are symmetric functions of the three squared momenta $x_i=\bar p_i^2$, and correspondingly of the $y_{ij}$. Structurally, we thus have
\begin{equation}
\begin{aligned}
 &\left[(6-2d) - 2y_{ij} \, \partial_{y_{ij}} - 2u \, \partial_u \right] \gamma_3(y_{12},y_{13},y_{23}|u) \\
 &= \gamma_3(y_{12},y_{13},y_{23}|u) \mathcal L_3(y_{12},y_{13},y_{23}|u) \\
 &+ \mathcal C_3(y_{12},y_{13},y_{23}|u) \, .
\end{aligned}
\end{equation}
The solution with the correct boundary condition at vanishing momenta is
\begin{equation}
\begin{aligned}
 &\gamma_3(y_{12},y_{13},y_{23}|u) \\
 &= -\frac{1}{2} \int_0^1 \text{d}\omega \, \omega^{d-4} \, \mathcal C_3(\omega y_{12},\omega y_{13}, \omega y_{23}|\omega u) \times \\
 &\qquad\qquad \exp \left[ -\frac{1}{2} \int_\omega^1 \frac{\text{d}\tau}{\tau} \mathcal L_3(\tau y_{12}, \tau y_{13}, \tau y_{23}|\tau u) \right] \, .
\end{aligned}
\end{equation}
The solution can be mapped to the coupling functions $W$ appearing in \eqref{eq:BMW-exp}, but the expression is rather lengthy and provides no additional insight, so we refrain from spelling it out.

Indeed, this structure carries over in the same way for the higher order correlation functions. The generalisation of the above for the $n$-point function, with $n\geq 3$, reads
\begin{equation}
\begin{aligned}
 &\left[(2n-(n-1)d) - 2y_{ij} \, \partial_{y_{ij}} - 2u \, \partial_u \right] \gamma_n(\{y_{ij}\}|u) \\
 &= \gamma_n(\{y_{ij}\}|u) \mathcal L_n(\{y_{ij}\}|u) + \mathcal C_n(\{y_{ij}\}|u) \, ,
\end{aligned}
\end{equation}
where
\begin{equation}
 \mathcal L_n = \sum_{i=1}^n \gamma_2(p_i^2|u) \mathcal I_{2,1}(p_i^2|u) \, ,
\end{equation}
comes from the self-energy type diagrams. All squared momenta should be replaced by combinations of the $y_{ij}$, and $\mathcal C_n$ arises from all flow diagrams which involve only correlators of order less than $n$, and thus are fixed already. The corresponding solution to the equation is
\begin{equation}
\begin{aligned}
 \gamma_n&(\{y_{ij}\}|u) = -\frac{1}{2} \int_0^1 \text{d}\omega \, \omega^{\frac{d-2}{2}(n-1) - 2} \, \mathcal C_n(\{\omega y_{ij}\}|\omega u) \times \\
 &\qquad\qquad\qquad\qquad \exp \left[ -\frac{1}{2} \int_\omega^1 \frac{\text{d}\tau}{\tau} \mathcal L_n(\{\tau y_{ij}\}|\tau u) \right] \, .
\end{aligned}
\end{equation}
Since the flow diagrams entering $\mathcal C_n$ can be calculated easily in an iterative way, this completes the calculation of the exact fixed point action for the $O(N)$ model at large $N$. We emphasise that while the general idea of an exact solution in the large $N$ limit has appeared in the literature before \cite{Blaizot:2005xy}, the systematic discussion of the fixed point correlation functions, filling in all intermediate details, has been missing.

\subsection{Regulator dependence}\label{sec:ONregdep}

We will now discuss the regulator dependence of the exact solution and its properties. Clearly, as it must be, the existence of the fixed point and all critical exponents are regulator-independent, see \eqref{eq:ONcritexp}. The latter are physically observable, at least in principle, and characterise the second order phase transition. Further regulator-independent features are the existence of a positive vacuum expectation value at $r=r_0>0$, the monotonicity of the derivative of the potential, and the bound on the value of $u(0)$. These are however qualitative, not quantitative features.

The more interesting question is thus about quantitative regulator independence, \eg{}, whether any combination of coupling constants is regulator-independent as well. If such a combination would be found, it would be possible to assess the quality of a given truncation in a new way. On the other hand, if this is not the case, so that coupling constants are essentially free and completely independent of each other, then any general comparison of values for coupling constants obtained with different regulators is meaningless. In particular, any strong regulator dependence of strictly non-universal quantities would not be a sign of a bad truncation whatsoever. We will find that indeed the latter option is the case: fixed point values of coupling constants are not universal.

To make this more precise, let us study the coefficients of the Taylor expansion of the critical potential around the minimum. Taking $n$ derivatives of the fixed point equation and using Fa\`a di Bruno's formula gives
\begin{equation}
\begin{aligned}
	&(2n-(n-1)d) v^{(n)}(r) - (d-2)r v^{(n+1)}(r) \\
	&= \frac{1}{2} \sum_{k=1}^n (-1)^k k! \, \mathcal I_{k+1}(v'(r)) B_{n,k}(v''(r),\dots,v^{(n-k+2)}(r)) \, .
\end{aligned}
\end{equation}
Here, the $B_{n,k}$ denote the Bell polynomials. We now notice that the only term on the right-hand side which involves $v^{(n+1)}(r)$ is the term with $k=1$. Evaluating this equation at the minimum $r=r_0$, we find that any dependence on $v^{(n+1)}(r_0)$ drops out of the equation, and we get a linear equation for $v^{(n)}(r_0)$ in terms of the lower derivatives at the minimum. The first few derivatives read
\begin{equation}
\begin{aligned}
	v(r_0) &= \frac{1}{2} \mathcal I_1(0) \, , \\
	v'(r_0) &= 0 \, , \\
	v''(r_0) &= \frac{4-d}{\mathcal I_3(0)} \, .
\end{aligned}
\end{equation}
In general, one finds that $v^{(n)}(r_0)$ depends on the threshold integral $\mathcal I_{n+1}(0)$. For $n\geq3$, the explicit dependence on this integral reads
\begin{equation}
	v^{(n)}(r_0) = \frac{1}{2} \frac{n!}{d-2n} \left( \frac{d-4}{\mathcal I_3(0)} \right)^n \mathcal I_{n+1}(0) + \dots \, ,
\end{equation}
where the dots indicate terms which only include threshold integrals with indices $\leq n$. Since these threshold integrals are in general independent of each other, we conclude that all couplings are independent of each other, and no (finite) combination of couplings can be found which is independent of the regulator choice.

In a similar way, one can perform an expansion around infinite field values, $r=\infty$. Here, one finds that successive expansion coefficients involve the threshold integrals with negative index, $\mathcal I_{-n}(0)$,\footnote{For this expansion to exist, we thus have to assume that the integrals
\begin{equation*}
	\int \text{d}\mu_q \dot R_k(q^2) \left(q^2 + R_k(q^2) \right)^n \, ,
\end{equation*}
are finite for all positive $n$. This is the case if the (scale derivative of the) regulator falls off faster than any polynomial, that is it is a Schwartz function on $\mathbbm R_+$.} which are in general independent of each other and of those with positive index. Thus there doesn't even seem to be a combination of terms from different expansions which combines to a regulator-independent quantity. We conclude that in general, nothing that doesn't need to be regulator-independent is regulator-independent.

\subsection{Beyond large \texorpdfstring{$N$}{N}}\label{sec:ONfiniteN}

\subsubsection{\texorpdfstring{$1/N$}{1/N} expansion}

An obvious question to ask is whether we can reproduce the results of the $1/N$ expansion. Unfortunately, this seems to be practically extremely hard, since already at order $1/N$, both the derivative expansion and its partial resummation fail to be perfect truncations. This can easily be seen by considering the exact flow equation for the potential -- at order $1/N$, it will always depend on the (field- and momentum-dependent) wave function renormalisation at the same order, and not only on the large $N$ solution. Likewise, the flow of the wave function renormalisation will depend on the higher order correlators of the same order in $1/N$. In a large $N$ expansion, the only simplification beyond the leading order is that the equations are linear, but in general they will form an infinite tower of integro-differential equations. This doesn't exclude that by clever rewriting, the corrections to the critical exponents and the anomalous dimension cannot be extracted, but we haven't found a way to do so. We conclude that to derive  exact expressions for higher order terms of the large $N$ expansion of critical quantities, standard large $N$ methods \cite{Moshe:2003xn, Klebanov:2018fzb} are more efficient.

\begin{figure}
\includegraphics[width=0.8\columnwidth]{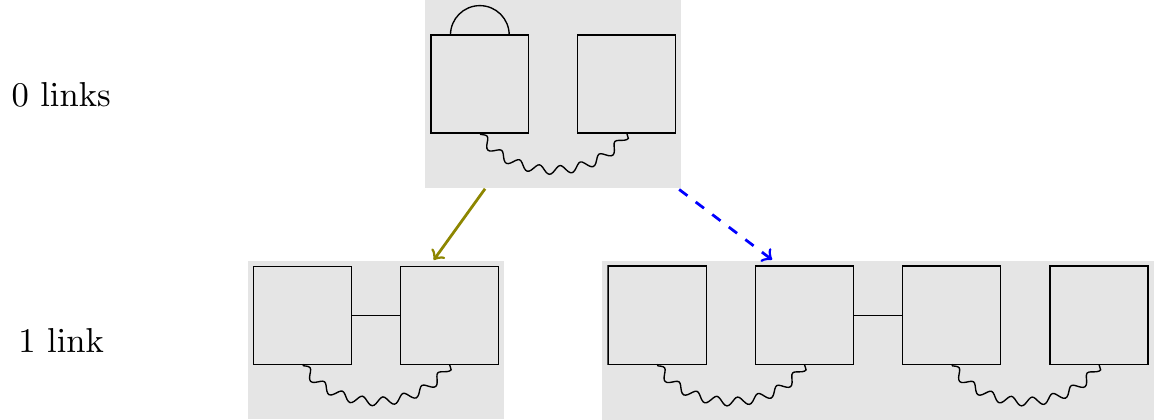}
\caption{The graph indicating the partial integration rules and the basis for the first order of the derivative expansion of the $O(N)$ model at finite $N$. In the convention of \cite{DePolsi:2020pjk}, the bottom row corresponds to the operators with prefactors $\frac{1}{2}Z$ and $\frac{1}{4}Y$, from left to right.}
\label{fig:d2_graph_finiteN}
\end{figure}

\begin{figure*}
	\includegraphics[width=\textwidth]{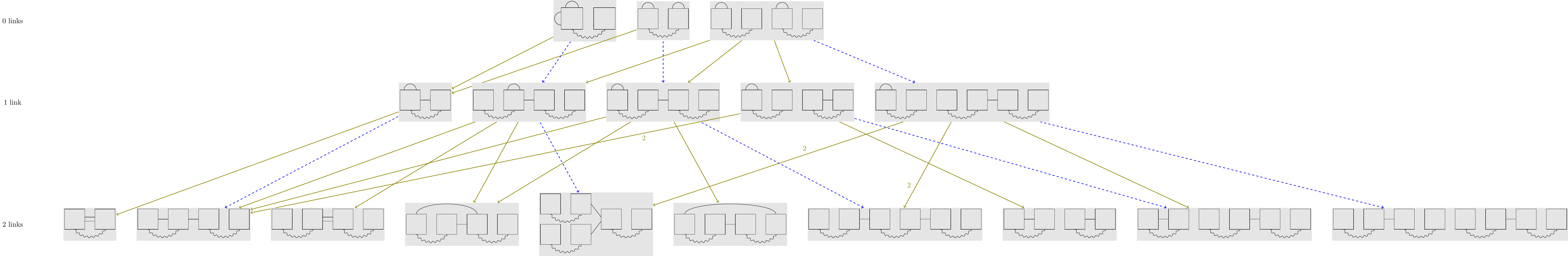}
	\caption{The graph indicating the partial integration rules and the basis for the second order of the derivative expansion of the $O(N)$ model at finite $N$. From left to right, the bottom row corresponds to the operators with prefactors $\frac{1}{2}W_1$, $W_3$, $\frac{1}{2}W_2$, $\frac{1}{2}W_4$, $\frac{1}{2}W_5$, $\frac{1}{4}W_7$, $\frac{1}{2}W_8$, $\frac{1}{4}W_6$, $\frac{1}{2}W_9$ and $\frac{1}{4}W_{10}$ of the investigation done in \cite{DePolsi:2020pjk}.}
	\label{fig:d4_graph_finiteN}
\end{figure*}

\subsubsection{Finite \texorpdfstring{$N$}{N}}

Let us now discuss some implications and benefits of the presented results for studies at finite $N$. From the analysis of the $O(N)$ model at varying $N$, it was found that the fixed point has a smooth dependence on $N$ \cite{DePolsi:2020pjk}. We thus propose the following new way to close the flow equation at finite $N$. Instead of setting correlation functions that are not resolved to zero, the large $N$ solution might be substituted instead. For large enough $N$, this should still be an excellent approximation, and improve the convergence of any other truncation scheme. This gives a practically feasible implementation of the idea presented in \cite{Kluth:2020bdv} to use information on higher order operators to stabilise lower order truncations.

On a conceptual level, at finite $N$ it is indeed more useful to employ a somewhat different basis than that of the large $N$ limit, since we now have to deal with a bundle index. It is most convenient to then let all derivatives act on fields $\phi^a$ only instead of both $\phi^a$ and $\rho$, since that way we can avoid having to introduce two kinds of boxes and deal with more complicated integration rules. This modifies our algorithm to obtain the basis and partial integration rules in the following way:
\begin{itemize}
 \item boxes now stand for fields $\phi^a$, and we have a second kind of link, which we represent by a wavy link, indicating contraction of the bundle index,
 \item only an even number of boxes can exist, since all fields must be contracted to form a scalar,
 \item when adding a new element by partial integration corresponding to a weight of $-x$, we have to add two boxes which are linked by a wavy link, and the derivative link connects to one of them; no factors of two arise because of the normalisation $\rho = \phi^a \phi^a/2$ so that $(\partial_\mu \rho) = \phi^a (\partial_\mu \phi^a)$.
\end{itemize}
We illustrate these changes in the graphs corresponding to the first two non-trivial orders in the derivative expansion in \autoref{fig:d2_graph_finiteN} and \autoref{fig:d4_graph_finiteN}. This choice of basis coincides with the choice made in the recent investigation of the $n=2$ truncation of the $O(N)$ model at finite $N$ \cite{DePolsi:2020pjk}, and we indicate the map to the prefactors of that work to our basis in the caption of the graphs. As has been noted in \cite{DePolsi:2020pjk}, for the case $N=1$ once again different rules apply since the wavy links don't represent a contraction anymore and can be dropped. Then, any pair of completely unconnected boxes can be reabsorbed into the prefactor coupling function, so that the number of independent operators at any given order of the derivative expansion is lower for $N=1$ than for $N>1$.

We will close this section by commenting again on the relation of this basis to the \BMW{} scheme \cite{Blaizot:2005xy, Blaizot:2005wd, Blaizot:2006vr, Benitez:2009xg, Benitez:2011xx}. At finite $N$ and order $n$, this corresponds to the resummation of diagram elements where exactly $n$ boxes are connected by at least one derivative link. In contrast to the large $N$ limit, this includes graphs with up to $2n$ boxes.

\section{Gross-Neveu model}\label{sec:GNmodel}

\subsection{Exact flow equation}\label{sec:GNflow}

As a second example for an exact fixed point solution, we will discuss the partially bosonised Gross-Neveu model. In this, we discuss $N_f$ Dirac fermions in a $d_\gamma$-dimensional representation of the Clifford algebra, coupled to a $\mathbbm Z_2$-symmetric scalar field. For investigations at finite $N_f$ in different dimensions and in both a condensed matter and a high energy physics context, see \cite{Gracey:1990sx, Gracey:1991vy, Luperini:1991sv, Rosenstein:1993zf, Karkkainen:1993ef, Gracey:1993kb, Gracey:1993kc, Braun:2010tt, Janssen:2014gea, Pawlowski:2014zaa, Braun:2014ata, Wang:2014cbw, Vacca:2015nta,  Borchardt:2015rxa, Knorr:2016sfs, Gracey:2016mio, Borchardt:2016xju, Mihaila:2017ble, Iliesiu:2017nrv, Zerf:2017zqi, Huffman:2017swn, Ihrig:2018hho, Dabelow:2019sty, Lenz:2020bxk, Lenz:2020cuv}. The large $N$ limit consists in taking the flavour number to infinity, $N_f\to\infty$, with the consequence that only diagrams with a complete fermion loop contribute to the flow in this limit. By assumption, the four-fermion coupling is set to zero in this model at the microscopic level since it has been transformed to a Yukawa coupling by the Hubbard-Stratonovich transformation. In that way, the fermionic sector of the action doesn't flow in the large $N$ limit, but remains classical. Employing again locality and regularity arguments, this fixes the form of the action to
\begin{equation}
 \Gamma_{N\to\infty}[\phi,\Psi] = d_\gamma N_f \left[ \int \text{d}^dx \left[ \mathbf{i} \bar \Psi \slashed\partial \Psi + \mathbf{i} \bar g \phi \bar \Psi \Psi \right] + \bar \Gamma[\phi] \right] \, .
\end{equation}
Both the fermionic wave function renormalisation and the Yukawa interaction $\bar g$ are restricted to be constant by locality and regularity. With the regulator choice
\begin{equation}
 \Delta S_k = d_\gamma N_f \int \text{d}^dx \, i \bar \Psi \slashed\partial \, r_k(-\partial^2) \Psi \, ,
\end{equation}
where $r_k$ is a dimensionless shape function, we can write down the exact flow as
\begin{widetext}
\begin{equation}\label{eq:GNflow}
 \dot{\bar\Gamma} = - \int \text{d}^dx \, \int \text{d}\mu_q \text{tr} \left[ (\mathbf{i} \slashed\partial - \slashed{q})(1+r_k((q-\mathbf{i}\partial)^2)) + \mathbf{i} \bar g \phi \right]^{-1} \left( -\slashed{q} \dot r_k(q^2) \right) \Bigg|_{d_\gamma N_f} \, ,
\end{equation}
\end{widetext}
where the remaining trace over the Dirac indices has to be projected onto the term proportional to $d_\gamma N_f$. To bring this into a more useful form, we first introduce the fermionic propagator at constant scalar field $\phi$,
\begin{equation}
\begin{aligned}
 G_0(q|\phi) &= \left[ - \slashed{q}(1+r_k(q^2)) + \mathbf{i} \bar g \phi \right]^{-1} \\
 &= - \frac{\slashed{q}(1+r_k(q^2)) + \mathbf{i} \bar g \phi}{q^2(1+r_k(q^2))^2 + 2\bar g^2\rho} \, .
\end{aligned}
\end{equation}
From the general flow \eqref{eq:GNflow} we can now extract the flows of all correlation functions at constant field, similar to the case of the $O(N)$ model. Since the only coupling between the bosonic and the fermionic sector is the field-independent Yukawa coupling, only diagrams with three-point functions will appear in the flow. In that way, we can write explicitly
\begin{widetext}
\begin{equation}
 \frac{\delta^n\dot{\bar \Gamma}}{\delta\phi^n}(p_1,\dots,p_n|\phi) = -(-\mathbf{i} \bar g)^n n! \int \text{d}\mu_q \, \text{tr} \, G_0(q|\phi) \slashed{q} \dot r_k(q^2) G_0(q|\phi) G_0(q+p_1|\phi) \cdots G_0(q+p_1+\dots+p_{n-1}|\phi) \Bigg|_{d_\gamma N_f} \, .
\end{equation}
\end{widetext}
A unit strength symmetrisation over the external momenta is implied on the right-hand side of the equation.

Let us point out two key differences to the $O(N)$ model: first, since the flow only contributes to bosonic operators but is driven purely by fermionic fluctuations, the resulting fixed point equations for the correlation functions are linear first order differential equations that can be integrated directly. Second, as is well-known, since the flow contributes to all bosonic operators, we will have a non-trivial bosonic anomalous dimension $\eta_\phi \neq 0$.

\subsection{Low order critical correlators}\label{sec:GNcorrelators}

We will now discuss the lowest order correlation functions and the spectrum. In the following we will occasionally use the identification
\begin{equation}\label{eq:regrewriting}
 q^2 (1+r_k(q^2))^2 = q^2 + R_k(q^2) \, ,
\end{equation}
to bring the flow equation into a form which resembles a bosonic flow.

Going somewhat out of order, we will first discuss the fixed point equation for the Yukawa coupling $\bar g$. The reason for this is that it is dimensionful, so that requiring a fixed point in fact determines the bosonic anomalous dimension. For this we assume that $\bar g\neq0$, which is reasonable since otherwise the flow is a constant, and we are only left with the Gaussian fixed point. In general dimensions, the flow equation for the dimensionless Yukawa coupling $g$ reads
\begin{equation}\label{eq:GNYukawaflow}
 \dot g = \frac{d-4+\eta_\phi}{2} g = 0 \, .
\end{equation}
We conclude that for $g\neq0$, the bosonic anomalous dimension is
\begin{equation}\label{eq:GNeta}
 \eta_\phi = 4-d \, .
\end{equation}

Next, we will discuss the bosonic potential. Its flow equation reads
\begin{equation}
\begin{aligned}
 \dot v(r) = -d v(r) + &(d-2+\eta_\phi) r v'(r) \\
 & - \frac{1}{2} \int \text{d}\mu_q \frac{\dot R_k(q^2)}{q^2 + R_k(q^2) + 2g^2 r} \, .
\end{aligned}
\end{equation}
Here we used the rewriting \eqref{eq:regrewriting}. Using \eqref{eq:GNeta}, the regular fixed point solution reads
\begin{widetext}
\begin{equation}\label{eq:GNFPpotential}
 v(r) = - \frac{1}{2d} \int \text{d}\mu_q \frac{\dot R_k(q^2)}{q^2+R_k(q^2)} \, {}_2F_1\left( 1, -\frac{d}{2}; 1-\frac{d}{2} \bigg | - \frac{2g^2 r}{q^2+R_k(q^2)} \right) \, .
\end{equation}
For non-negative $r$ the potential is monotonically increasing, so that the fixed point potential is in the symmetric regime. Let us now discuss the bosonic two-point function. At constant $r$, the flow equation reads\footnote{We stress that in this section, $\gamma_2$ denotes the second variation of the effective action with respect to $\phi$, in contrast to the discussion on the $O(N)$ model, where the variation was with respect to $\rho$.}
\begin{equation}\label{eq:GN2ptflow}
\begin{aligned}
 \dot \gamma_2(p^2|r) &= (-2+\eta_\phi) \gamma_2(p^2|r) + (d-2+\eta_\phi) r \, \gamma_2^{(0,1)}(p^2|r) + 2p^2 \gamma_2^{(1,0)}(p^2|r) \\
 & \qquad+ 2 g^2 \int \text{d}\mu_q \frac{\dot r_k(q^2)}{\left[ q^2(1+r_k(q^2))^2 + 2g^2 r \right]^2 \left[ (p+q)^2(1+r_k((p+q)^2))^2 + 2g^2 r \right]} \times \\
 &\qquad\qquad\qquad\qquad \bigg\{ (q^2 + p\cdot q)(1+r_k((p+q)^2))(q^2(1+r_k(q^2))^2-2g^2 r) - 4q^2(1+r_k(q^2))g^2 r \bigg\} \, .
\end{aligned}
\end{equation}
\end{widetext}
Denoting the integral (without the prefactor) as $\mathcal I^\psi_2(p^2|r)$,\footnote{Here we again used that by Lorentz invariance, the integral must be a function of $p^2$.} the solution to the equation with the correct boundary conditions reads
\begin{equation}\label{eq:GN2ptsol}
\begin{aligned}
	\gamma_2(p^2|r) &= \frac{2g^2}{d-2} \mathcal I_2^\psi(0|0) \\
	&-g^2 \int_0^1 \text{d}\omega \, \omega^{-\frac{d}{2}} \left[ I^\psi_2(\omega p^2| \omega r) - I^\psi_2(0 | 0) \right] \, .
\end{aligned}
\end{equation}
One can easily check that this reduces to $v'(r)+2rv''(r)$ at $p=0$, as it must for consistency, and that
\begin{equation}
 \gamma_2(p^2|r) \propto \left( p^2 \right)^{\frac{d}{2}-1} = \left( p^2 \right)^{1-\frac{\eta_\phi}{2}} \, , \qquad \text{as } p\to\infty \, ,
\end{equation}
which is consistent with the scaling expected from the anomalous dimension \eqref{eq:GNeta}.

It is worthwhile to discuss the equation for the anomalous dimension, which is related to the canonical normalisation of the scalar field. For this, we take the $p^2$-derivative of \eqref{eq:GN2ptflow} and set $p=r=0$. Using the normalisation condition $\gamma^{(1,0)}(0,0)=1$ to fix the wave function renormalisation, we find
\begin{equation}\label{eq:GNetaeq}
\begin{aligned}
	\frac{\eta_\phi}{g^2} &= \frac{1-\frac{1}{d}}{2} \int \text{d}\mu_q \frac{\dot R_k(q^2)}{q^2(q^2+R_k(q^2))^3} \\
	&\quad + \frac{3}{2} \int \text{d}\mu_q \frac{\dot R_k(q^2) \left( 1+R_k'(q^2) + \frac{2}{d}q^2 \, R_k''(q^2) \right)}{(q^2+R_k(q^2))^4} \\
	&\quad -\frac{15}{2d} \int \text{d}\mu_q q^2 \, \frac{\dot R_k(q^2)(1+R_k'(q^2))^2}{(q^2+R_k(q^2))^5} \, .
\end{aligned}
\end{equation}
While usually this equation fixes the anomalous dimension, in our case it actually fixes the value of the Yukawa coupling $g$, since $\eta_\phi$ is already fixed by \eqref{eq:GNeta}. This equation will be important in the next subsection to discuss the spectrum of the theory.

The fixed point equations for the higher order correlation functions can be found in a similar way, and solutions have a similar form as \eqref{eq:GN2ptsol}. We note that for all correlators, including the two-point function, general solutions are only defined up to a multiple of the homogeneous solution, but this is however uniquely fixed by the consistency condition at vanishing momenta and regularity. We thus will not present the higher order correlators explicitly.

\subsection{Critical exponents}\label{sec:GNcritexp}

We are now prepared to discuss the critical exponents of the fixed point. First, we will discuss the spectrum that is related to variations of the potential, with vanishing variation of the Yukawa coupling and anomalous dimension. With these conditions, we find
\begin{equation}
	-\theta \delta v(r) = -d \delta v(r) + (d-2+\eta_\phi) r \delta v'(r) \, ,
\end{equation}
with the solution
\begin{equation}
	\delta v(r) = c \, r^{\frac{d-\theta}{2}} \, .
\end{equation}
Here, $c$ is a constant. In fact, these are the same perturbations as for the $O(N)$ model, with the replacement $r\to u$, once again reinforcing that for the $O(N)$ model $u$ is the more natural variable. In conclusion, this part of the spectrum agrees with the spectrum of the $O(N)$ model,
\begin{equation}\label{eq:GNtheta}
	\theta = d-2n \, , \qquad n \in \mathbbm{N} \, .
\end{equation}
There is one additional critical exponent which corresponds to a finite perturbation of the Yukawa coupling. Taking the variation of \eqref{eq:GNYukawaflow}, we get
\begin{equation}\label{eq:GNdeltag}
	-\theta \delta g = \frac{1}{2} g \, \delta \eta_\phi \, ,
\end{equation}
where we used the fixed point value for the anomalous dimension. On the other hand, the variation of \eqref{eq:GNetaeq} also relates the variations of $\eta_\phi$ and $g$ via
\begin{equation}
	\delta \left( \frac{\eta_\phi}{g^2} \right) = 0 \, ,
\end{equation}
so that
\begin{equation}\label{eq:GNdeltaeta}
	\delta \eta_\phi = \frac{2\eta_\phi}{g} \delta g \, .
\end{equation}
Combining \eqref{eq:GNdeltag} with \eqref{eq:GNdeltaeta}, we find the final critical exponent
\begin{equation}\label{eq:GNthetaeta}
	\theta_\eta = -\eta_\phi = d-4 \, .
\end{equation}
Notice that due to the linear nature of the fixed point equations, it is clear that one can find solutions to the perturbation equations of all $n$-point correlation functions for all critical exponents. For example, for $\delta g=0$, we find
\begin{equation}
	(d-\theta-2) \delta \gamma_2(p^2|r) = 2 r \, \delta \gamma_2^{(0,1)}(p^2|r) + 2p^2 \delta \gamma_2^{(1,0)}(p^2|r) \, .
\end{equation}
The general solution to this equation reads
\begin{equation}
	\delta \gamma_2(p^2|r) = \left(p^2\right)^{\frac{d-2-\theta}{2}} c_n\left( \frac{r}{p^2} \right) \, ,
\end{equation}
where $c_n$ is a free function. The set of critical exponents \eqref{eq:GNtheta} then makes the first factor regular, and requiring that the perturbation is well-defined for all momenta constrains $c_n$ to be a polynomial of order at most $(n-1)$. Since $\delta\gamma_2(0|r)$ is related to $\delta v(r)$, there will be conditions on certain coefficients of this polynomial in relation to the normalisation constant $c$ of the perturbations of the potential.

It is interesting to point out explicitly that almost all critical exponents of the $O(N)$ model and the Gross-Neveu model at large $N$ agree. The only exception is the extra critical exponent $\theta_\eta$ which is exclusive to the latter model. This has the interesting consequence that to uniquely characterise any universality class, the \emph{complete} set of critical exponents has to be determined. Any finite or infinite subset is, in general, insufficient to specify the universality class.

\subsection{Regulator dependence}\label{sec:GNregdep}

Once again we discuss the regulator dependence of the solution. The existence of the fixed point and its spectrum are independent of the regulator, as required. By contrast, the situation for the coupling constants is the exact same as for the $O(N)$ model. Again taking the example of the couplings making up the potential, the Taylor coefficients can be directly read off from the functional solution \eqref{eq:GNFPpotential}, since the hypergeometric function can be defined in terms of a power series around vanishing argument. In complete similarity to the $O(N)$ model, the $n$-th coupling constant is related to an independent threshold integral, $\mathcal I_n(0)$. We conclude that all remarks made earlier also apply to the Gross-Neveu model, and indicate that indeed generically, fixed point couplings carry no regulator-independent information at all.

\section{Summary}\label{sec:summary}

In this work we studied the $O(N)$ model and the Gross-Neveu model within functional renormalisation in dimensions between two and four. In both models, we derived the complete non-perturbative fixed point effective action in the large $N$ limit without specifying the regulator function. That was possible because of the simplification of the flow equation in the large $N$ limit which allows for perfect truncations, where improving the truncation doesn't alter the renormalisation group equations of previous orders.

For the $O(N)$ model, with regards to a large $N$ expansion, we have found indications that already at next-to-leading order, the perfect truncation property of the large $N$ limit is lost. It thus seems difficult to derive the $1/N$ corrections to critical quantities from the functional renormalisation group analytically. Nevertheless, the form of the effective action at infinite $N$ might serve as an approximate closure of the flow equation at finite $N$, potentially improving the convergence of common approximation schemes.

The Gross-Neveu model is even simpler than the $O(N)$ model in the large $N$ limit, since the flow is exclusively driven by fermionic fluctuations contributing to the running of bosonic operators. This results in linear partial differential equations for the correlation functions. Notably, the spectrum of the Gross-Neveu model and the $O(N)$ model is almost the same.

A key result of the study of these models is that only observables and qualitative aspects are regulator-independent. Concretely, this concerns the existence of the fixed point and its spectrum, and qualitative results like the existence or absence of a non-trivial vacuum expectation value. By contrast, generically fixed point couplings depend on the choice of the regulator even for non-truncated solutions to the flow equation, and thus are essentially arbitrary. This is not entirely surprising, but having concrete calculations without systematic errors supporting this view is reassuring.

To our knowledge, this is the first time that a complete non-perturbative fixed point action was derived. This serves as a showcase to test general properties of the effective action independent of truncation errors. The explicit fixed point correlators represent a useful testing bed for numerical methods that aim to resolve field and momentum dependencies.

As a new technical result, we have put forward an algorithm to derive an operator basis at $n$-th order of the derivative expansion, which has an intuitive graphical representation, and that at the same time provides the partial integration rules to map any operator of a given order into the basis. Its application to higher orders, while cumbersome, is entirely straightforward, and the algorithm can be implemented in a computer code to automatise these computations.

It would be interesting to derive the conformal data \cite{Wilson:1969zs, Codello:2017hhh, Pagani:2020ejb}, such as the coefficients of the operator product expansion, for these analytically solvable cases to gain some knowledge about what techniques work best to extract the data from a non-perturbative renormalisation group flow, but we leave this for future work.

\section*{Acknowledgements}
This work was supported by Perimeter Institute for Theoretical Physics. Research at Perimeter Institute is supported in part by the Government of Canada through the Department of Innovation, Science and Industry Canada and by the Province of Ontario through the Ministry of Colleges and Universities.

\bibliography{gen_bib}

\end{document}